\documentclass{article}
\usepackage[dvips]{graphics}
\usepackage{array,hhline,amssymb,amsthm,color,amsfonts}
\def\L{{\mathfrak{L}}}
\def\O{{\mathcal{O}}}  \def\o{{\mathbf{o}}}
\def\K{{\overline{\O}}}  \def\k{{\overline{\o}}}
   
\def\R{\mathbb{R}}  \def\C{\mathbb{C}}    
\def\d12{\raise0.4ex\hbox{\scriptsize 1}\!/\!\lower0.4ex\hbox{\scriptsize 2}}
\def\tr{{\rm tr}} 
\def\bux#1#2{\buildrel{*}\over{#1}_{#2}}
\title{Model of the electro-weak, gravitational and strong interactions in the O-theory}
\date{}
\author{V. Yu. Dorofeev\thanks{Dep. of Math., SPb SUEF,
Sadovaya 21, 191023, St.Petersburg,
Russia, E-mail: friedlab@mail.ru}\\
Friedmann Laboratory for Theoretical Physics}
\begin{document}
\maketitle
\begin{abstract}
Based on the matrix representation of octonion algebra, supplied
with specific multiplication rule, the model of electroweak and
gravitational interactions is built up. While electroweak
interaction in this model is induced by charged W-bosons, other two forces appear to have slightly more complicated
nature. Gravitational interaction coincides in the model with dipole
interaction of a pair of charged bosons. The dipole consists of a charged vector
bosons pair from the major octonion algebra fields. When the charged dipole
pair interacts with the neutral bosons pair from the major octonion algebra fields, the charged bosons pair misses its mass. The drop in mass leads to appearance of far-ranging
forces of gravitational interaction. Finally, strong interaction
appears in the model as internal gravitational solution of
'black whole' type with the peculiar 'gravitational' constant. The
solution is a product of interaction of major vector fields pair
with charged W-bosons pair. It is inferred from the
model that the state space is ten-dimensional. The space
is built as a module of the matrix representation of octonion
algebra over the particles field (O-module). Similarly to
the Standard Weinberg-Salam theory, the particle mass here appears
as the product of interaction of massless spinor fields and Higgs
field from O-module representation.
\end{abstract}
\section*{Introduction}
The idea of incorporation of octonion algebra into total field
framework is a long-life story \cite{Jordan}. However, the early
attempts were mostly concerned in the new Cayley octaves algebra
introduction to the physics theory necessity rather than in seeking
approach for unified field theory construction. Nowadays octonions
are included in the emerging M-theory framework of unified
interactions. Nevertheless the group approach to M-theory construction
has a sensitive disadvantage: Cayley octaves algebras lose
non-associativity of their lagrangian.

In the current work principally different approach to unified field
theory construction is proposed. This approach implies the algebraic
introduction of new interactions on the special algebra. Such
approach has a number of advantages, though along with certain
disadvantages. The main disadvantage is a necessity of new
quantization schemes introduction. That becomes inevitable because
the lagrangian loses its important group properties when considered
on the newly introduced algebra. Therefore the problem of
probabilistic interpretation of particles and the positivity of its
energy arises. However after some reasoning that fact does not face
so negative. As it follows from calculations, new particles should
have higher mass therefore their influence within the modern energy
scale could be neglected without losing any of group properties of
$U(1)\times SU(2)$ symmetries of electro-weak interaction. In case
of high energies non-associative nature of the lagrangian provides
us with one specific opportunity: the order of multipliers we could
choose according to our own preferences. That opportunity allows us
to exclude undesirable summands. Thus in the proposed
fields interaction model new quantization as well as renormalization
methods, perhaps, could be found.

In \cite{Dor1} the author proposed generalization of the lagrangian
from Weinberg-Salam theory to non-associative algebra and later in
\cite{Dor2} proposed the scheme of gravitational interaction
introduction. The main milestones of conducted research would be
point out in the first paragraphs of the paper. After that strong
interaction introduction scheme as a model of strong gravity, but in
the framework of the approach developed here, and as a solution inside of
``black whole'' in the Schwarzschild metric, is proposed. New gravitation
constant of strong gravitation is defined by the mass of charged
$W$-bosons from weak interaction theory.

\section{Dirac equation\\ in the pseudo-riemannian space}
Let $M$ be some pseudo-riemannian manifold and the coordinates $x(p)=(x^0,x^1,$ $x^2,x^3)=x$ may be introduced in any point $p\in M$, metric is 
\begin{equation}\label{p1_1}
ds^2=g_{\mu\nu}dx^\mu dx^\nu
\end{equation}
and the connectness is
\begin{equation}\label{p1_2}
\Gamma^\mu_{\nu\lambda}=\frac12g^{\mu\kappa}(g_{\mu\kappa,\nu}+
g_{\nu\kappa,\lambda}-g_{\lambda\nu,\kappa})
\end{equation}

Then the Riemannian tensor we define by
\begin{equation}\label{p1_3}
R^\tau_{\mu\nu\lambda}=\Gamma^\tau_{\mu\lambda,\nu}-
\Gamma^\tau_{\mu\nu,\lambda}+\Gamma^\tau_{\sigma\nu}\Gamma^\sigma_{\mu\lambda}-
\Gamma^\tau_{\sigma\lambda}\Gamma^\sigma_{\mu\nu}.
\end{equation}

Quadratic form (\ref{p1_1}) may be reduced to the diagonal form in any neighborhood of the point
\begin{equation}\label{p1_4}
ds^2=H_0^2dx^{(0)2}-
H_1^2dx^{(1)2}-H_2^2dx^{(2)2}-H_3^2dx^{(3)2}
\end{equation}

We consider the metric of the physical space-time as metric in Minkowsky space $M_4$ in the normal form
\begin{equation}\label{p1_5}
ds^2=c^2dt^2-dx^2-dy^2-dz^2=\eta_{ab}dx^adx^b
\end{equation}
(Here we make distinction between the Greek indexes and Latin indexes: the Greek indexes are used with regard to pseudo-riemannien manifold and the Latin indexes are used with regard to the $M_4$.)

Build the tangent space in every point of the pseudo-riemannian space $M$. Assume the tangent space is the Minkowsky space $M_4$ with metric (\ref{p1_5}). Let 
\begin{equation}\label{p1_6}
H_0dx^{(0)}=cdt,H_1dx^{(1)}=dx,H_2dx^{(2)}=dy,H_3dx^{(3)}=dz
\end{equation}

Introduce the tetrads $h^a_\mu$ in the same points of the pseudo-riemannian space $M$ to connect the metric of pseudo-riemannian space to the metric of the $M_4$:
\begin{equation}\label{p1_7}
h^b_\mu h^\mu_\nu=\delta^b_a,\qquad h^{\mu(a)}h^\nu_a=g^{\mu\nu} \end{equation}

Let $h_\mu^a$ be tetrads \cite{Landau}: $A^a=A^\mu h^a_\mu$. Then
\begin{equation}\label{p1_8}
\delta A^\mu=\delta(A^ah^\mu_a)=\delta A^ah^\mu_a+A^a\delta h^\mu_a=\delta A^ah^\mu_a+A^ah^\mu_{a,\nu}\delta x^\nu=\Gamma^\mu_{\nu\lambda}A^\nu\delta x^\lambda \end{equation}

As $h^b_\mu h^\mu_{a,\nu}+h^b_{\mu,\nu} h^\mu_a=0$ so
\begin{equation}\label{p1_9}
\delta A^b=\gamma^b_{ac}A^a\delta x^c,\qquad\gamma^b_{ac}=
h^b_{\mu;\nu}h^\mu_ah^\nu_c
\end{equation}
where $\gamma^b_{ac}$ are Ricci coefficients.

In \cite{FokIvanenko} free Dirac equation
\begin{equation}\label{p1_10}
(i\gamma^a\partial_a-m)e(x)=0
\end{equation}
in the tangential space $M_4$ of the pseudo-riemanienn space have the following form
\begin{equation}\label{p1_11}
(ih^\mu_a\gamma^a(\partial_\mu-i\Phi_\mu^1-\Gamma_\mu)-m)e(x)=0
\end{equation}
where
\begin{equation}\label{p1_12}
\Gamma_a=h_a^\mu\Gamma_\mu=-\frac12\gamma_{abc}\sigma^{bc},\qquad \sigma^{bc}=\frac14[\gamma^b,\gamma^c],\qquad a,b=0,1,2,3
\end{equation}
and $\Phi_\mu^1$ is arbitrary real function.

Dirac matrixes comply with the following multiplication rule on matrixes $\sigma^{ab}$:
\begin{equation}\label{p1_13}
\gamma^a\sigma^{bc}=\frac14\gamma^a[\gamma^b,\gamma^c]=\frac12\eta^{ab}\gamma^c-\frac12\eta^{ac}\gamma^b-\frac i2\varepsilon^{dabc}\gamma^5\gamma_d
\end{equation}

Then
\begin{equation}\label{p1_14}
-\gamma^a\Gamma_a=
\frac14h^\mu_ah^\nu_bh_{(c)\nu;\mu}(\frac12\eta^{ab}\gamma^c-\frac12\eta^{ac}\gamma^b-\frac i2\varepsilon^{dabc}\gamma^5\gamma_d)
\end{equation}

Assume, the metric (\ref{p1_1}) has form (\ref{p1_4}), then for different $a,b,c$ we get 
\begin{equation}\label{p1_15}
\Gamma^\lambda_{\mu\nu}h_{c\lambda}h^\mu_bh^\nu_c=\frac12g^{\mu\kappa}(g_{\mu\kappa,\nu}+
g_{\nu\kappa,\lambda}-g_{\lambda\nu,\kappa})h_{c\lambda}h^\mu_bh^\nu_c=0
\end{equation}
so
\begin{equation}\label{p1_16}
-\gamma^a\Gamma_a=\frac14h^\mu_ah^{\nu(a)}h_{(c)\nu;\mu}\gamma^c-\frac14
h^{\mu(a)}h^\nu_bh_{(a)\nu;\mu}\gamma^b$$
$$=\frac14h^\mu_{c;\mu}\gamma^c+\frac14h^{\mu(a)}h^\nu_{b;\mu}h_{(a)\nu}\gamma^b=\frac12h^\mu_{c;\mu}\gamma^c
\end{equation}
and Dirac equation has the form \cite{SokolovIvanenko}
\begin{equation}\label{p1_17}
(i\gamma^ah^\mu_a(\partial_\mu-i\Phi_\mu^1+\Phi_\mu^2)-m)\psi=0
\end{equation}
where
\begin{equation}\label{p1_18}
\Phi_\mu^2=\frac12\partial_\mu\left(\ln\frac{\sqrt{-g}}{H_\mu}\right)
\end{equation}

\section{$\K$ - space}
The doubling of quaternion algebra leads, in particular, to
octonion  algebra $\tilde\O$ \cite{Baez}, which is a linear space over the field of real numbers $\R$, for any $\tilde\o$ from octonion algebra $\tilde\O$. Octonion algebra cannot be represented by matrixes with traditional multiplication rule, but the special multiplication rule
can be introduced, which allows such representation \cite{DD}
\begin{equation}\label{p2_1}
\tilde\O=\{\forall\tilde\o=\sum_{A=0}^7\alpha^A\tilde\Sigma^A=\alpha^A\tilde\Sigma^A,\quad\alpha^A\in\R\}
\end{equation}
where $\tilde\Sigma^K$ are matrix and ($I,J,K=1,2,\dots,7$):
\begin{equation}\label{p2_2}
\tilde\Sigma^I\cdot\tilde\Sigma^J=-\delta^{IJ}+\varepsilon^{IJK}\tilde\Sigma^K,
\end{equation}
where completely antisymmetric symbol $\varepsilon^{IJK}$ is not
null if only
$$\varepsilon^{123}=\varepsilon^{145}=\varepsilon^{176}=\varepsilon^{246}=
\varepsilon^{257}=\varepsilon^{347}=\varepsilon^{365}=1$$
and  $\tilde\Sigma^0=1$ (matrice $2\times2$.

It is easy to ensure, that the multiplication rule leads to
non-associative algebra, i.e.
$$\{\tilde\Sigma^A,\tilde\Sigma^B,\tilde\Sigma^C\}=(\tilde\Sigma^A\tilde\Sigma^B)\tilde\Sigma^C-
\tilde\Sigma^A(\tilde\Sigma^B\tilde\Sigma^C)=2\varepsilon^{ABCD}\tilde\Sigma^D$$
where $\varepsilon^{IJKL}$ is completely antisymmetric symbol, which
is equal to one for the following expressions:
\begin{equation}\label{p2_3}
\varepsilon^{1247}\;=\;\varepsilon^{1265}\;=\;\varepsilon^{2345}
\;=\;\varepsilon^{2376}\;=\;\varepsilon^{3146}\;=\;\varepsilon^{3157}
\;=\;\varepsilon^{4567}\;=1
\end{equation}

Let $\Sigma^K=i\tilde\Sigma^K,K=1,2,$ $\dots,7$ and
\begin{equation}\label{p2_4}
\O=\{\forall\o=\alpha^A\Sigma^A,\quad\alpha^A\in\C,A=0,1,\dots,7\}
\end{equation}
where ($I=1,2,3$)
\begin{equation}\label{p2_5}
\matrix{\Sigma^0=\left(\matrix{\sigma^0&0\cr0&\sigma^0}\right)&\Sigma^I=
\left(\matrix{0&-i\sigma^I\cr i\sigma^I&0}\right)\cr
\Sigma^4=\left(\matrix{-\sigma^0&0\cr0&\sigma^0}\right)&
\Sigma^{4+I}=\left(\matrix{0&-\sigma^I\cr -\sigma^I&0}\right)}
\end{equation}
Here $\sigma^0$ is a unit matrix and $\sigma^I,I=1,2,3$ is a Pauli matrix:
\begin{equation}\label{p_2_6}
\sigma^1=\left(\matrix{0&1\cr1&0}\right),\quad
\sigma^2=\left(\matrix{0&-i\cr i&0}\right),\quad
\sigma^3=\left(\matrix{1&0\cr0&-1}\right),\quad
\sigma^0=\left(\matrix{1&0\cr0&1}\right)
\end{equation}

Let's define the specific multiplication rule for the abstract matrixes in the
following way
$$\o*\o'=\left(\matrix{\lambda I&A\cr B&\xi I}\right)*
\left(\matrix{\lambda' I&A'\cr B'&\xi' I}\right)=$$
\begin{equation}\label{p2_7}
=\left(\matrix{(\lambda\lambda'+\frac12\tr(AB'))I\hfill&\lambda
A'+\xi' A+\frac i2[B,B']\hfill\cr\lambda'B+\xi B'-\frac
i2[A,A']&(\xi\xi'+\frac12\tr(BA'))I\hfill}\right)
\end{equation}

The matrixes $A,A',B,B$ of size $(2\times2)$ were introduced, together with unit matrix $I$ of the same size and complex numbers $\lambda,\xi,\lambda',\xi'$.

Multiplication rule (\ref{p2_7}) is defined for a wider matrix class than  octonion algebra representation matrixes $\Sigma^a,a=0,1,2,\dots,7$ introduced earlier. Two more matrixes could be introduced
\begin{equation}\label{p2_8}
f^8=\left(\matrix{0&I\cr
I&0}\right),\quad f^9=\left(\matrix{0& iI\cr-iI&0}\right)
\end{equation}
That maintains our specific multiplication rule (\ref{p2_7}).

So the set of matrixes
\begin{equation}\label{p2_9}
f^a=\Sigma^a,\quad a=0,1,\dots,7,\quad f^8,\quad f^9
\end{equation}
form a basis of linear space over the field of complex numbers, which is further called the extended octonion space, and is denoted for $\K$.

That defines the space $\K$ as non-associative $\O$-module, where $\O$ is alternative ring.

By means of multiplication (\ref{p2_7}) introduce the convolution on the $\K$:
\begin{equation}\label{p2_11}
(\k_1,\k_2)=\frac12\tr(\k_1^+*\k_2 )=\frac12\tr(\left(\matrix{\lambda_1^*I&B_1^+\cr A_1^+&\xi_1^*I}\right)*
\left(\matrix{\lambda_2I&A_2\cr B_2&\xi_2I}\right))$$
$$=\lambda_1^*\lambda_2+\xi_1^*\xi_2+\frac12\tr(B_1^+B_2))
+\frac12\tr(A_1^+A_2)
\end{equation}
where $\K\times\K$ in $\C$. 

\section{Methods of relieving of octonions\\ from non-associativity}
{\bf Probability model of non-associativity relieving.}
\begin{equation}\label{p3_1}
A*B*C*D=p_1((A*B)*C)*D+A*(B*C)*D
\end{equation}
$$p_1+p_2=1$$

Here the frequencies $p_1$ and $p_2$ are introduced. These values depend on the number of the brackets permutations, which lead to the equal result in (\ref{p3_1}). For example, in (\ref{p3_1}) it is natural to admit $c_1^{ab}=p_1=3/4,c_2^{ab}p_2=1/4$ because three different types of brackets permutations lead to the equal result in the first member of the right side of the equation (\ref{p3_1}) whereas only one type in the second member of the right side of (\ref{p3_1}).

The definition given in (\ref{p3_1}) is also applicable for greater number of elements.

{\bf Minimal model of non-associativity relieving.}
\begin{equation}\label{p3_2}
A*B*C*D=\min\{(A*B)*(C*D),A*(B*C)*D\}
\end{equation}

{\bf The maximal or minimax model} is to be defined similarly.

Obviously, in associative case the same result is obtained to that of the models defined above.

\section{Lagrangian of the O-theory}
Let 
\begin{equation}\label{p4_1}
{\bf A}_b(x)=A_b^A(x)\Sigma^A,\quad \Sigma^A\in\O_\R,A=0,1,2,\dots,7
\end{equation}
$$\O_\R=\{\forall\o_\R=\alpha^A\Sigma^A,\quad\alpha^A\in\R,A=0,1,\dots,7\}$$
where $A_b^A(x)\in\R,x\in M_4,b=0,1,2,3$.

In \cite{Dor1} the generalization of Weinberg-Salam lagrangian to non-associative algebra is proposed as follows ($A,B=0,1,\dots,7$):
$$\L_\o=\L_f+(\partial_a\bux\Psi\varphi-
\frac i2q^AA_a^A\bux\Psi\varphi*\Sigma^A)
*(\partial^a\Psi_\varphi+
\frac i2q^BA^{a(B)} \Sigma^B*\Psi_\varphi)$$
$$+\frac i2\overline L*\gamma_a(\overrightarrow\partial^a
+\frac i2c_Aq^AA^{a(A)}\Sigma^A)*L
-\frac i2\overline L*\gamma_a(\overleftarrow\partial^a-
\frac i2c_Aq^AA^{a(A)}\Sigma^A)*L$$
$$+\frac i2\overline R\gamma_a*(\overrightarrow\partial^a R+
iq^0A^{a0}R)-\frac i2\overline
R\gamma_a*(\overleftarrow\partial^a R-iq^0A^{a0}R)$$
\begin{equation}\label{p4_2}
-\tilde h\overline L*\Psi_\varphi*R-\tilde h\overline
R*\bux\Psi\varphi*L)+m^2||\Psi_\varphi||^2-\frac f4||\Psi_\varphi||^4
\end{equation}

Here $\Sigma^A\in\O,\quad\Psi_\varphi,\Psi\in\K$.

The lagrangian of the free fields $\L_f$ is ($I,J,K,L=1,\dots,7$)
\begin{equation}\label{p4_3}
\L_f=-\frac14F^0_{ab}F^{ab(0)}-\frac1{16}\tr(\hat F_{ab}*\hat  F^{ab})=-\frac14F^0_{ab}F^{ab(0)}-\frac1{4}F_{ab}^KF^{ab(K)}$$ $$-\frac14f^{IJKL}q^{IJ}q^{KL}(A_a^IA_b^J-A_b^I
A_a^J)(A^{a(K)}A^{b(L)}-A^{b(K)}A^{a(L)})
\end{equation}
\begin{equation}\label{p4_4}
f^{IJKL}=\frac14\tr(\Sigma^I*\Sigma^J*\Sigma^K*\Sigma^l)
\end{equation}
$$\hat F_{ab}=\partial_b A_a^K\Sigma^K-\partial_a  A_b^K\Sigma^K+q^{IJ}(A_a^IA_b^J-A_b^IA_a^J)\Sigma^I*\Sigma^J$$
$$F^0_{ab}=\partial_b A^0_a-\partial_a A^0_b,\qquad
F_{ab}^k=\partial_\nu A_a^k-\partial_a A_b^k+\varepsilon^{IJK}q^{IJ}(A_a^IA_b^J-A_b^IA_a^J)$$

Left and right spinor components
\begin{equation}\label{4_5}
\frac12(1+\gamma^5)\Psi=L,\qquad\frac12(1-\gamma^5)\Psi=R,\qquad\gamma^5=i\gamma^0\gamma^1\gamma^2\gamma^3
\end{equation}

$\Psi,\Psi_\varphi\in\K$. $c_L,c_{L_0}$ are some numbers, caused by normalization; $q^A,q^{JK}$ are charges; $\gamma^a$ are Dirac matrixes ($a=0,1,2,3;i=1,2,3$):
\begin{equation}\label{p4_6}
\gamma^0=\left(\matrix{I&0\cr0&-I}\right),\qquad
\gamma^i=\left(\matrix{0&\sigma^i\cr-\sigma^i&0}\right),\qquad \gamma^5=\left(\matrix{0&I\cr I&0}\right)
\end{equation}

The summing up by different-case Greek indices is carried out with metric tensor of Minkowsky space $\eta_{\mu\nu}$ with signature $(1,-1,-1,-1)$, while the simple summing up is conducted by the same-case indices.

It is shown in \cite{Dor1} ($R=e_R(x),\overline L=L^+\gamma^0$),
\begin{equation}\label{p4_7}
L=\frac{c_0}{\sqrt2}\left(\matrix{0\qquad(2i\sigma^1-2\sigma^2)\nu(x)
+(y_0I+\frac{2i}3\sigma^1+\frac23\sigma^2+i\sigma^3)e(x)\cr(-\frac
i8\sigma^1+\frac18\sigma^2)\nu(x)+(y_0I-\frac{3i}8\sigma^1-\frac38\sigma^2+
\frac{9i}{16}\sigma^3)e(x)\qquad0}\right)_L
\end{equation}
if
\begin{equation}\label{p4_8}
\Psi_\varphi=\Psi_0=\frac{m}{\sqrt{2f}}
\left(\matrix{0&i\sigma^3\cr0&I}\right)
\end{equation}
the lagrangian of the O-theory takes the following form for the lepton sector
$$\L_{\o}=\L_f+
\frac{q^{(K)2}m^2}{2f}A_a^KA^{a(K)}+\frac{o^{IJ}m^2}{2f}A_a^{i}A^{a(j)}+\frac{g^{(1)2}m^2}{2f}B_a B^a-\frac{gg^{(1)}m^2}fA_a^3B^a$$
$$+\frac{g^{(1)}}2\overline\nu_L\gamma_a B^a\nu_L+
\frac{g^{(1)}}2\overline e_L\gamma_aB^a e_L+\frac g2\overline
e_L\gamma_aA^{a3}e_L-\frac g2\overline\nu_L\gamma_aA^{a3}\nu_L$$
$$-\frac g2\overline\nu_L\gamma_ae_L(A^{a1}-iA^{a2})-
\frac g2\overline e_L\gamma_a\nu_L(A^{a1}+iA^{a2})$$
$$+\frac i2(\overline e_L\gamma_a\partial^a e_L-\partial^a\overline
e_L\gamma_a e_L)+\frac i2(\overline\nu_L\gamma_a\partial^a\nu_L-
\partial^a\overline\nu_L\gamma_a\nu_L)+\frac{m^4}f$$
$$+\frac i2(\overline e_R\gamma_a\partial^a
e_R-\partial^a\overline e_R\gamma_a e_R)+g^{(1)}\overline
e_R\gamma_a B^a e_R-\frac{\sqrt2hm}{\sqrt f}(\overline
e_Le_R+\overline e_Re_L)$$
$$-q^4A^{a(4)}(\kappa_1\overline\nu_L\gamma_a\nu_L- \kappa_2\overline
e_L\gamma_ae_L)-\frac32q^6A^{a(6)}\overline e_L\gamma_ae_L$$
\begin{equation}\label{p4_9}
-\frac54(q^6A^{a(6)}+iq^5A^{a(5)})\overline\nu_L\gamma_a
e_L-\frac54(q^6A^{a(6)}-iq^5A^{a(5)})\overline e_L\gamma_a\nu_L
\end{equation}

Where the following notation is used $ic_2^{ij}A^{ij}q^iq^j=o^{ij},\kappa_1,\kappa_2\approx10$. 

So, the final lagrangian contains non-associative elements along with associative.

Consider non-associative summands from lagrangian $\L_o$. First of all, it is the quadratic member by the fields $A_\mu^k$ 
\begin{equation}\label{p4_10}
\frac{o^{IJ}m^2}{2f}A_a^{I}A^{a(J)}
\end{equation}

Apply probability model of non-associativity relieving to ths member. Assume for non-zero components $c^{ij}_1=c^{ji}_1=3/4, c^{ij}_2=c^{ji}_2=1/4$. Hence due to the symmetry of the expression $q^iq^jA_\mu^{i}A^{\mu(j)}$ with respect to $i,j$ (there is no summing up by them) and the anti-symmetry of multiplier $A^{ij}_k$ (again with respect to $i,j$) this member equals zero.

In addition there is a non-associative term (\ref{p4_4}) in the lagrangian of free fields. Relieving from non-associativity for this member will lead to important physical properties of the lagrangian of the O-theory, therefore the discussion of methods of relieving it from non-assotiativity postpone to the appropriate section.

Assume the senior field $A_a^K,K=4,5,6,7$ equal zero, we come to the lagrangian of the Weinberg-Salam, the Standard Theory of weak interactions (ST)  \cite{Okun} in the appropriate gauge. Strictly speaking, the problem formulation concludes in choosing such generalization to $\K$-module that in the particular case of the minor fields we would get exactly the lagrangian of the ST.

\section{Octonion lagrangian research}
1. Current fields of $A_\mu^k,k=0,1,2,3$, as it is in the ST, indicate the presence of the vector bosons $Z_0,W$ and $W^*$, which appeared to be massive after their quadratic part of lagrangian and vacuum, induced by $\Psi_0$, research. This fact is expected, since the general principle of deducing the O-theory lagrangian concludes in obtaining such lagrangian $\L_\o$ that, after excluding of the major fields $A_\mu^k,k=4,5,6,7$, we get the Lagrangian of the ST.

2. By analogy to the ST, the current containing part
\begin{equation}\label{p5_1}
q^4A^{\mu(4)}(\kappa_2\overline e_L\gamma_\mu e_L-\kappa_1\overline\nu_L\gamma_\mu\nu_L)
\end{equation}
of the lagrangian (\ref{p4_9}) $\L_\o$ gives the opportunoty to introduce a neutral vector field $C_\mu=A_\mu^4$. Also, the form of quadratic terms of field $A_\mu^4$ allows us to introduce the lagrangian of a massive vector field $C_\mu$
\begin{equation}\label{p5_2}
\L_C=-\frac14(\partial_\mu C_\nu-\partial_\nu C_\mu)
(\partial^\mu C^\nu-\partial^\nu C^\mu)+\frac{m^2_C}2C_\mu C^\mu,\quad
m^2_C=\frac{(q^{(4)})^2m^2}f
\end{equation}

3. The field $A_\mu^7$ is special, because there are no currents. Research of the quadratic terms $A_\mu^7$ gives foundation to introduce the massive vector field $E_\mu=A_\mu^7$ and the lagrangian
\begin{equation}\label{p5_3}
\L_E=-\frac14(\partial_\mu E_\nu-\partial_\nu E_\mu)
(\partial^\mu E^\nu-\partial^\nu E^\mu)+\frac{m^2_C}2E_\mu E^\mu,\quad
m^2_E=\frac{(q^{(7)})^2m^2}f
\end{equation}

4. The current containing part
\begin{equation}\label{p5_4}
\frac54(q^6A^{\mu(6)}+iq^5A^{\mu(5)})\overline\nu_L\gamma_\mu
e_L+\frac54(q^6A^{\mu(6)}-iq^5A^{\mu(5)})\overline e_L\gamma_\mu\nu_L
\end{equation}
gives foundation to introduce the charged vector field
\begin{equation}\label{p5_5}
D_\mu=\frac1{2q_D}(q^6A^{\mu(6)}-iq^5A^{\mu(5)})
\end{equation}
and the quadratic terms of the lagrangian $\L_\o$ allows to introduce a massive vector field $D_\mu$ with lagrangian (where $m_D^2=2m^2q_D^2/f$)
\begin{equation}\label{p5_6}
\L_D=-\frac12(\partial_\mu D_\nu^*-\partial_\nu D_\mu^*)
(\partial^\mu D^\nu-\partial^\nu D^\mu)+m_D^2D_\mu^*D^\mu
\end{equation}

However, that must be assumed $q^{(5)2}=q^{(6)2}=q_D^2=q_{D^*}^2$, thereforee $m_5=m_6=m_D=m_{D^*}$.

But not everything is so smooth! There is another interesting term in the lagrangian $\L_\o$
\begin{equation}\label{p5_7}
-\frac32q^6A^{\mu(6)}\overline e_L\gamma_\mu e_L
\end{equation}

In a way, this term is also responsible for the current. Indeed
\begin{equation}\label{p5_8}
-\frac32q^6A^{\mu(6)}\overline e_L\gamma_\mu e_L=
-\frac34(q_{D^*}D_\mu^*+q_DD_\mu)\overline e_L\gamma^\mu e_L
\end{equation}
but this kind of the current violates the invariance of the lagrangian under the global transformation of the charge for vector fields: if
\begin{equation}\label{p5_9}
D_\mu\to e^{iQ}D_\mu,\quad e_L\to e^{iQ}e_L\quad,\hbox{ý. ñ.}
\end{equation}
then
\begin{equation}\label{p5_10}
(q_{D^*}D_\mu^*+q_DD_\mu)\overline e_Le_L\to (e^{-iQ}q_{D^*}D_\mu^*+e^{iQ}q_DD_\mu)e^{-iQ}\overline e_Le^{iQ}e_L$$
$$\ne(q_{D^*}D_\mu^*+q_DD_\mu)\overline e_Le_L
\end{equation}

Loss of global charge invariance is due to the appearance of a massive vector boson, with respect to a non-zero vacuum value $\Psi_0$. However, the unusual form of the current containing term (\ref{p5_8}) requires further study. 

\section{Solutions for vector bosons in the O-theory}
We write the Euler-Lagrange equations for vector fields $A_\mu^k,k=5,6$ lagrangian $\L_\o$. Let
$$F_{\mu\nu}^5=A_{\nu,\mu}^5-A_{\mu,\nu}^5,\quad
F_{\mu\nu}^6=A_{\nu,\mu}^6-A_{\mu,\nu}^6$$
then we get
\begin{equation}\label{p6_1}
F^{\mu(5)}_{\nu,\mu}+m_5^2A_\nu^5-\frac14f^{IJK5}q^{K5}q^{IJ}A^{\mu(I)}A_\nu^KA_\mu^J=0
\end{equation}
\begin{equation}\label{p6_2}
F^{\mu(6)}_{\nu,\mu}+m_6^2A_\nu^6-\frac14f^{I6KJ}q^{J6}q^{IK}A^{\mu(I)}A_\nu^JA_\mu^K=0
\end{equation}
where $m_i=m_D,i=5,6$. (In the equations (\ref{p6_1} \ref{p6_2}) there is non-abelian part of the lagrangian fields $\L_f$, which is there omitted.)

Part of the lagrangian, reflecting its non-associative nature, remained in $f^{IJKL}$. Non-zero values of $f^{IJKL}$ in the $\L_\o$ are equal to $\pm1$ only for some fields permutations (\ref{p2_2}). Choose $f^{IJKL}=-1$ according to minimax model. If the only non-zero $f^{4675}=-1$ in (\ref{p6_1} -- \ref{p6_2}) we get
\begin{equation}\label{p6_3}
F^{\mu(5)}_{\nu,\mu}+m^2_DA_\nu^5-q^{47}q^{56}A_\mu^4A_\nu^6A_\mu^7=0
\end{equation}
\begin{equation}\label{p6_4}
F^{\mu(6)}_{\nu,\mu}+m^2_DA_\nu^6-q^{47}q^{56}A_\mu^4A_\nu^5A_\mu^7=0
\end{equation}
convolution product $A_\mu^4A^{\mu(7)}$ is a scalar and we shall believe it to be a constant on the small region $\Omega_i$. We also assume that all values in (\ref{p6_3} -- \ref{p6_4}) are chosen so that
$$m_D^2=q^{47}q^{56}A^{\mu(4)}A_\mu^7$$
in particular,
\begin{equation}\label{p6_5}
A^{\mu(4)}A_\mu^7=\frac{m_D^2}{q^{47}q^{56}}
\end{equation}

However, then there is a massless solution of the equations (\ref{p6_3} -- \ref{p6_4}) for the vector-potentials $A_\nu^5$ è $A_\nu^6$:
\begin{equation}\label{p6_6}
A_\nu^5=A_\nu^6
\end{equation}
of the form (recall (\ref{p6_5}))
\begin{equation}\label{p6_7}
F^{\mu\nu(D)}_{,\mu}=0,\quad F_{\nu\mu}^D=\partial_\mu D_\nu-\partial_\nu D_\mu
\end{equation}
Thus the $D$-boson is massless. As a solution of the equation (\ref{p6_7}) in the spherical coordinate system it takes
\begin{equation}\label{p6_8}
A^5_\mu=A^6_\mu=(0,f(r),0,0)
\end{equation}
or
\begin{equation}\label{p6_9}
A^5_\mu=A^6_\mu=(f(t),0,0,0)
\end{equation}
where $f(r)$ and $f(t)$ are arbitrary functions of the radius vector $r$ and time $t$, respectively.

On the other side
\begin{equation}\label{p6_10}
F^{\mu\nu(C)}_{,\mu}+m_C^2C^\nu=q^{47}q^{56}A^5_\mu A^{\mu(6)}E^\nu$$
$$F^{\mu\nu(E)}_{,\mu}+m_E^2E^\nu=q^{47}q^{56}A^5_\mu A^{\mu(6)}C^\nu
\end{equation}

Until now, the fields were considered solely as classic. When calculating the convolution $C_\mu E^\mu$ it is not sufficient. Let the convolutionà $C_\mu E^\mu$ be an average
\begin{equation}\label{p6_11}
C_\mu E^\mu=<|\hat C_\mu\hat E^\mu|>\ne0
\end{equation}
where
\begin{equation}\label{p6_12}
\hat C_\mu=l_\mu^s(\hat c_s^+e^{-ikx}+\hat c_s e^{ikx}),\
\hat E^\mu=r_\mu^s(\hat e_s^+e^{-ikx}+\hat e_se^{ikx}),\ k^2=m^2
\end{equation}
also we introduced creation and annihilation operators of Bose-particles and assumed the same mass for particles $C$ and $E$. Since the particles $C,E$ are different and the operator $\hat c,\hat e$ are commutative, then the right part of the (\ref{p6_11}) by the vacuum state is equal to zero. At the same time the (\ref{p6_11}) by the bound state is not equal to zero. For example, assume $|1,1>=\hat c^+\hat e^+)|0>$ then
\begin{equation}\label{p6_13}
<0|(\hat c+\hat c^+)(\hat e^++\hat e)\hat c^+\hat e^+|0>=
<0|\hat c\hat c^+\hat e\hat e^+|0>\ne0$$
$$<0|\hat c\hat e(\hat c+\hat c^+)(\hat e^++\hat e)|0>=
<0|\hat c\hat c^+\hat e\hat e^+|0>\ne0
\end{equation}

Since the left side of (\ref{p6_10}) is zero  (recall $\partial_\mu C^\mu=\partial_\mu E^\mu=0$) \cite{Shirkov}, then $A_\mu^5A^{\mu(6)}$ should be required to be small.

\section{The method of geometrization in the O-theory}
As it was mentioned, the lagrangian $\L_\o$ includes the current vector, which is non-invariant under the transformation of charge (\ref{p5_10}). The reason is that the currents
\begin{equation}\label{p7_1}
q_DD_a^*(y)\overline e_L(y)\gamma^ae_L(y) \hbox{ è }
q_{D^*}D_a(y)\overline e_L(y)\gamma^ae_L(y)
\end{equation}
are included in the lagrangian as a sum. (Here $y^a,a=0,1,2,3$ are the variables of the flat Minkowsky space.) Moreover, by substituting this Lagrangian in the action we find that basically these terms are defined in distinct points of $U$
$$S'=\frac i2\int_U(\overline e\gamma^a(\partial_a+\frac{3i}2q_DD_a)e
-(\partial^a-\frac{3i}2q_{D^*}D^*_a\overline e\gamma_ae))d^4y$$
\begin{equation}\label{p7_2}
=\frac i2\int_U(\overline e\gamma^a(\partial_a+\frac{3i}2q_DD_a)ed^4y
-\frac i2\int_U(\partial_a-\frac{3i}2q_{D^*}D^*_a)\overline e\gamma^aed^4y
\end{equation}

(The differences between the left and right particles were neglected in the
paper, despite the fact that basically the considered space was Riemannian-Cartan
space. Detailed calcula\-tions show the matrix $\gamma^5$ leads
to Riemannian-Cartan space \cite{Dor3}, however, on the earth surface
the effect is insignificant and unobservable in the modern empirical
works (\cite{Russell}-\cite{Heck})).

On the other side in the initial lagrangian we had $A^6_\mu(y)$. Therefore pair $D$ and $D^*$, also in action, should be defined at one point. To solve this problem, perform the procedure of geometrization.

Let's limit by the case of ``sufficiently good connected $\Omega$'' and ``suffiently well-defined $\L_o$''.

Consider a sufficiently large region of Minkowsky space $\overline U=T\times\R^3$ where  $T$ and $\R^3$ are compacts. Split the domain $\overline U$ in an arbitrary way to $n$ different small compact domains  $\overline U_i\subset\overline U$ with boundary $\partial\overline U_i$ and with inner region $U_i=\overline U_i\backslash\partial\overline U_i$
\begin{equation}\label{p7_3}
\bigcup_i\overline U_i=\overline U,\quad U_i\bigcap_{i\ne j}U_j=\varnothing,\quad i,j=1,\dots,n.
\end{equation}

In the region $U_i$ arbitrary choose the points $A_i$ and introduce local coordinates $\xi_i$:
\begin{equation}\label{p7_5}
y_i=y(A_i)+\xi_i
\end{equation}

Then the interval between any two points from any neighborhood $U_i$ has form
\begin{equation}\label{p7_6}
ds^2=(dy^0)^2-(dy^1)^2-(dy^2)^2-(dy^3)^2
\end{equation}

Introduce manifold $\overline M$ and split it to $n$ regions $\overline M_i$ 
\begin{equation}\label{p7_7}
\bigcup_i\overline M_i=\overline M,\quad M_i\bigcap_{i\ne j} M_j=\varnothing,\quad i,j=1,\dots,n.
\end{equation}

Define homeomorphisms for the domains $M_i,\Omega_i$ è $U_i$
\begin{equation}\label{p7_70}
f_i:M_i\to\Omega_i\subset\R^4,$$
$$g_i:\Omega_i\to U_i,\quad p_i=f_i^{-1}\circ g_i^{-1}(A_i)
\end{equation}

Denote $x=x(p)=(x_0,\dots,x_3),p\in M, x_\mu(p)$ the coordinates of the $M$, which are induced by $f_i$. Assume $g_i,f_i$ and coordinate systems in $U_i$ and in $\Omega_i$ satisfy
\begin{equation}\label{p7_8}
y_a-y_a(p_i)=H_a^\mu(p_i)(x_\mu-x_\mu(p_i))
\end{equation}
where $H^\mu_a$ is a diagonal matrix.

Then the quadratic form of the interval with respect to $p_i$ from the manifold $M_i$ has the form:
\begin{equation}\label{p7_9}
ds^2=H_0^2(dx^0)^2-H_1^2(dx^1)^2-H_2^2(dx^2)^2-H_3^2(dx^3)^2
\end{equation}

Write the action of the O-theory lagrangian in the form of the Riemannian integral
\begin{equation}\label{p7_10}
S_\o=\int_U
\L_od^4y=\lim_{n\to\infty}\sum_{i=1}^n\L_o(A_i)
\Delta U_i,\quad\Delta U_i=\Delta y_0\Delta y_1\Delta y_2\Delta y_3
\end{equation}
or
\begin{equation}\label{p7_11}
S_\o=\lim_{n\to\infty}\sum_{i=1}^n\L_o(p_i)\sqrt{-g(p_i)}\Delta^4x,\quad \sqrt{-g(p_i)}=H_0H_1H_2H_3
\end{equation}

Rewrite the spinor part of the action
$$\Delta S_i'=(\frac i2(\overline e\gamma_\mu(H_\mu^{-1}\partial^\mu+\frac{3i}4(q^6A^{\mu(6)}-iq^5A^{\mu(5)}))e$$ $$-(H_\mu^{-1}\partial^\mu-\frac{3i}4(q^6A^{\mu(6)}+iq^5A^{\mu(5)})\overline e\gamma_ae))$$
\begin{equation}\label{p7_12}
-\frac58(q^6A^{\mu(6)}+iq^5A^{\mu(5)})\overline\nu\gamma_\mu
e-\frac58(q^6A^{\mu(6)}-iq^5A^{\mu5)})\overline e\gamma_\mu\nu) \sqrt{-g(p_i)}\Delta^4x
\end{equation}

Using the definition of vector fields $D$ and $D^*$ from (\ref{p5_5}) we come to the following expression for (\ref{p7_12})
$$\Delta S_i'=(\frac i2(\overline e\gamma^\mu(\partial_\mu+\frac{3i}2q_DD_\mu)e
-(\partial^\mu-\frac{3i}2q_{D^*}D^*_\mu\overline e)\gamma_\mu e))$$
\begin{equation}\label{p7_13}
-\frac54q_{D^*}D^*_\mu\overline\nu\gamma_\mu
e-\frac54q_DD_\mu\overline e\gamma_\mu\nu)\Delta\Omega_i,\quad \Delta\Omega_i=\sqrt{-g(p_i)}\Delta^4x
\end{equation}

From now on we neglect the decay of $D$-bosons into leptons, and therefore excluded the second string in (\ref{p7_13}) from further consideration.

In a small neighborhood $U_i$ of the point of Minkowsky space we consider the derivative of the spinor Dirac particle as a covariant derivative in curved space, written in the tetrad formalism in the metric (\ref{p1_4}), for which tetrads $H^\mu_a(p_i)$ are chosen in such way that
\begin{equation}\label{p7_14}
\Phi_\mu^2(p_i)=\frac34q^5A_\mu^5(p_i)\end{equation}
where $\Phi_\mu^2$ is same as (\ref{p1_18}).

In the new coordinates the action (\ref{p7_13}) has the following form
\begin{equation}\label{p7_15}
\Delta S_i'=(\frac i2(\overline e\gamma^\mu(\partial_\mu +\frac34q^5A^{\mu(5)}+\frac{3i}4q^6A^{\mu(6)})e$$ $$-(\partial^\mu+\frac34q^5A^{\mu(5)}- \frac{3i}4q^6A^{\mu(6)}\overline e\gamma_\mu e)) \Delta\Omega_i
\end{equation}
which is equal to
\begin{equation}\label{p7_16}
\Delta S_i'=(\frac i2(\overline e\gamma^\mu(\nabla_\mu+\frac{3i}4q_{D^*}A^6_\mu)e-(\nabla_\mu-\frac{3i}4q_DA^6_\mu)\overline e\gamma^\mu e))\Delta\Omega_i
\end{equation}
or
\begin{equation}\label{p7_17}
\Delta S_i'=(\frac i2(\overline e\gamma^\mu\nabla_\mu e-\nabla_\mu\overline e\gamma^\mu\nabla_\mu e+\frac{3i}4(q_{D^*}A^6_\mu e+q_DA^6_\mu)\overline e\gamma^\mu e))\Delta\Omega_i
\end{equation}

Since $A_\mu^6$ is given in the only point $p_i$ and the signs of the interaction of particles $D,D^*$ and electron are opposite, then
\begin{equation}\label{p7_18}
\Delta S_i'=\frac i2(\overline e\gamma^\mu\nabla_\mu e-\nabla_\mu\overline e\gamma^\mu e)\Delta\Omega_i
\end{equation}

Assuming that the integrand is extendable to the whole $M$ we obtain
\begin{equation}\label{p7_19}
S'=\int_\Omega\frac i2(\overline e\gamma^\mu\nabla_\mu e-\nabla_\mu\overline e\gamma^\mu e)\sqrt{-g}dx^4
\end{equation}

The written out scheme coincides to the representation of the Riemannian integral for an ordinary problem of the GRG, so the above assumptions are feasible, because they are complied for some problems. For example, such scheme is suitable for the calculation of Tolman \cite{Tolman} and Friedmann (\ref{p8_1}) solutions.

In addition it is necessary to take in mind the GRG provides the link between matter and geometry through Einstein equations. It turns out, the equations connecting matter and geometry exist in the Lagrangian of the O-theory. Indeed, the lagrangian of the O-theory includes an unusual member
\begin{equation}\label{p7_20}
\L_f'=\frac14f^{IJKL}q^{IJ}q^{KL}(A_a^IA_b^J-A_b^I
A_a^J)(A^{a(K)}A^{b(L)}-A^{b(K)}A^{a(L)})$$
$$=\tilde f^{IJKL}A_a^IA_b^JA^{a(K)}A^{b(L)}
\end{equation}
which sign is not defined before specifying the method of non-associativity relieving.

Small regions $\Omega_i$ differ in the following way: there are regions of $\Omega_i$, which imply the existence of a virtual pair of $D+D^*$-bosons, and the regions where there is no such pair. In the areas where there is a virtual pair of $D+D^*$ it will be assumed a virtual pair of $C\cdot E$-bosons also exists (the contribution of the minor fields $A_\mu^B,B=0,1,2,3$ will be neglected since their mass compared to that of major field $A_\mu^B,B=4,5,6,7$ is negligible quantity).

Outside of the region of the implied virtual pair of $D+D^*$-bosons the gravitational vacuum is not formed, hence the average value of a virtual pair $C+E$ in such region is zero.

Action of the gravitational field in GRG has the following form \cite{Landau}:
\begin{equation}\label{p7_21}
S_g=-\frac1\kappa\int_\Omega R\sqrt{-g}d\Omega=-\frac1\kappa\int_\Omega G\sqrt{-g}d\Omega- \frac1\kappa\int_\Omega\partial_\lambda(\sqrt{-g}w^\lambda)d\Omega
\end{equation}
where $w^\lambda$ is vector \cite{Landau} and
\begin{equation}\label{p7_22}
L_g=-\frac1\kappa G=-\frac1\kappa g^{\mu\nu}(\Gamma^\lambda_{\sigma\nu} \Gamma^\sigma_{\mu\lambda}-\Gamma^\lambda_{\sigma\lambda}\Gamma^\sigma_{\mu\nu})
\end{equation}
is lagrangian of the gravitational field.

We will not make distinction between $G$ and $R$ in the integrand because the divergance terms could omitted.

From here on, we assume (the validity of this assumption is verified in the cases of the Schwarzchild and Friedmann metric)
\begin{equation}\label{p7_23}
-\frac1\kappa\int_\Omega G\sqrt{-g}d\Omega=f^{4567}m_D^2\int_\Omega A_\mu^5A^{\mu(6)}\sqrt{-g}d\Omega
\end{equation}
(here $f^{4567}=0,\pm1$) and we come to Einstein equations
\begin{equation}\label{p7_25}
R_{\mu\nu}=\kappa T_{\mu\nu}
\end{equation}
where $\kappa=m_D^2$.

\section{Friedmann solution}
In this section it is shown in the flat Friedmannian space there is a self-consistent solution of octonion Lagrangian.

Consider homogeneous and isotropic Universe
\begin{equation}\label{p8_1}
ds^2=dx^{(0)2}-a^2(t)(dx^{(1)2}+dx^{(2)2}+dx^{(3)2})=
a^2(\eta)(d\eta^2-dl^2),
\end{equation}
with conformal time $dt=a(\eta)d\eta$ and ($\alpha,\beta=1,2,3$)
\begin{equation}\label{p8_2}
g_{00}=a^2(\eta),\quad g_{\alpha\beta}=a^2(\eta)\eta_{\alpha\beta}
\end{equation}

Use (\ref{p1_2}) to find all the non-zero components of Christoffel symbol:
\begin{equation}\label{p8_3}
\Gamma^0_{00}=\frac{a'}{a^3},\quad\Gamma^0_{\alpha\beta}=-\frac{a'}{a^3}g_{\alpha\beta},\quad\Gamma^\alpha_{0\beta}=\frac{a'}a\delta^\alpha_\beta
\end{equation}
and estimate the value
\begin{equation}\label{p8_4}
G=g^{\mu\nu}(\Gamma^\lambda_{\mu\nu}\Gamma^\kappa_{\lambda\kappa}- \Gamma^\lambda_{\mu\kappa}\Gamma^\kappa_{\nu\lambda})
=\frac{6{a'}^2}{a^4},
\end{equation}
where the stroke means the derivative with respect to conformal time.

Write the Dirac equation in the Friedmann metric (\ref{p8_1})
\begin{equation}\label{p8_5}
(\gamma^\mu(H^\mu)^{-1}(\partial_\mu-i\Phi_j+\frac12
\partial_\mu\left(\ln\frac{\sqrt{-g}}{H^\mu}\right))-m)\psi=0
\end{equation}

From (\ref{p1_18}) and (\ref{p8_5}) we get 
\begin{equation}\label{p8_6}
q_DA_\mu^5=(2da/(a^2d\eta),\vec 0)
\end{equation}
which is inline with (\ref{p6_9}).

Assume 
\begin{equation}\label{p8_7}
f^{4567}q^{56}q^{47}A^{(4)\mu}A^{(7)}_\mu=-\frac{3q^2_D}{2\kappa}
\end{equation}
then Lagrangian member (\ref{p4_4}), which is non-associative by fields, could be rewritten, using (\ref{p7_14}), as
\begin{equation}\label{p8_8}
f^{4567}q^{56}q^{47}A^{(5)\mu}A^{(6)}_\mu A^{(4)\nu}A^{(7)}_\nu=\frac{q^{56}q^{47}}{q_D^2}f^{4567}A^{(4)\mu}A^{(7)}_\mu
\left(\frac{2da}{a^2d\eta}\right)^2$$
$$=-\frac6{\kappa}\left(\frac{da}{a^2d\eta}\right)^2
=-\frac6{\kappa}\frac{a'^2}{a^4}=-\frac1{\kappa}G=\L_g
\end{equation}
(calculated based on the minimax non-associativity relieving method) and determines the gravitational field lagrangian.

Consider the right side of (\ref{p6_10}) equations of $C$ and $E$ bosons.
\begin{equation}\label{p8_15}
F^{\mu\nu(C)}_{,\mu}+m_C^2C^\nu=
q^{56}q^{47}\frac{4a'^2}{q_D^2a^4}E^\nu$$
$$F^{\mu\nu(E)}_{,\mu}+m_E^2E^\nu=
q^{56}q^{47}\frac{4a'^2}{q_D^2a^4}C^\nu
\end{equation}

If we assume that the mass of $C$ and $E$-bosons is large and the rate of Hubble recession is small, then, at the certain choice of constant $4q^{56}q^{47}/q_D^2$, the right side of (\ref{p8_15}) could indeed be regarded as negligibly small.

\section{Schwarzchild solution}
Consider the space where there is a massive spherically symmetrical object. Let this object be the source of octonion field. On great distances from the object the electro-weak interaction could be neglected therefore the object should be  the source of solely the major octonion fields. Due to symmetry of the problem we can take on the great distances all the above fields are produced by vector-potential $A_\mu^k=A_\mu^k(r),k=4,5,6,7$.

Let, e.g. the electron, be moving in the space. The space where the electron is moving is by definition the Minkowsky space. In spherically symmetrical coordinates the metrics in the space forms
\begin{equation}\label{p9_1}
ds^2=dt^2-dr^2-r^2(\sin^2\theta d\varphi^2+d\theta^2)
\end{equation}

Let's limit the consideration by left spinors. Rewrite the motion equation in octonion field of massive source, assumed it does not interact with the fields $A^{4,7}_\mu$
\begin{equation}\label{p9_2}
(i\gamma^0\partial_0+i\gamma^r(\partial_r-\vec\Sigma\cdot\hat{\vec L}-\frac{3i}4q^6A^6_r-\frac34q^5A^5_r)+m)\psi=0
\end{equation}
with denotation
$$\gamma^r=\gamma^1\sin\theta\cos\varphi+
\gamma^2\sin\theta\sin\varphi+\gamma^3\cos\theta,\quad
\vec\Sigma=\left(\matrix{\vec\sigma&0\cr0&\vec\sigma}\right)$$
where $\vec L=\vec r\times\vec p$ is the angular moment operator \cite{Wheeler1}.

In accordance with the general scheme of geometrization of the O-theory we get
\begin{equation}\label{p9_3}
\frac12\partial_r\left(\ln\frac{\sqrt{-g}}{H^r}\right))=-\frac34q^5A^5_r
\end{equation}

Introduce the general form of stationary spherically symmetric metric, obtained as the product of geometrization of the O-theory
\begin{equation}\label{p9_4}
ds^2=H_0^2(r)dt^2-H^2_1(r)dr^2-r^2(\sin^2\theta d\varphi^2+d\theta^2)
\end{equation}
therefore
\begin{equation}\label{p9_5}
-\frac34q^5A^5_r=\frac12\partial_r\left(\ln\frac{\sqrt{-\bar g}}{H^r}\right))=\frac{H_{0,r}}{2H_0},\qquad A^5=(0,A^5_r,0,0)
\end{equation}

For a weak field it is well-known  \cite{Landau}  
\begin{equation}\label{p9_6}
H_0^2=g_{00}=1-r_g/r=f^2
\end{equation}
where $r_g$ is gravitational radius.

So assume $H_0$ to be known and find $H_1(r)$. In the first approximation, with great $r$, assume $H_1=1+C/r^n,n>1$. 
\begin{equation}\label{p9_7}
H_1=1+C/r^n=1+\beta(r),\quad n>1
\end{equation}

Find gravitational field lagrangian (\ref{p7_13}) in metrics (\ref{p9_4})
\begin{equation}\label{p9_8}
G_g=g^{\mu\nu}(\Gamma^\lambda_{\mu\nu}\Gamma^\kappa_{\lambda\kappa}- \Gamma^\lambda_{\mu\kappa}\Gamma^\kappa_{\nu\lambda})
=\frac2{r^2H_1^2}+\frac{4H_{0,r}}{rH_1^2H_0}-\frac{H_{0,r}H_{1,r}}{H_1^3H_0}
\end{equation}

Since the gravitational field (\ref{p7_13}) is $2/r^2$ for the flat metric (\ref{p9_1}) then the lagrangian of the gravitational field in the E-theory, as it follows from (\ref{p7_14}) and (\ref{p9_5}), is
\begin{equation}\label{p9_9}
G_g=G_{pl}+C_0q^{(5)2}A_\mu^5A^{(5)\mu}=\frac2{r^2}+C_0\frac{4H_{0,r}^2}{9H_0^2}
\end{equation}

Given $\alpha(r),\beta(r)$ are infinitesimal, we see that matching (\ref{p9_8}) and (\ref{p9_9}) can only be achieved provided that
\begin{equation}\label{p9_10}
H_0H_1=1,\qquad g_{11}=(1-r_g/r)^{-1}
\end{equation}

Finally the spherically-symmetrical Schwarzchild metric is obtained
\begin{equation}\label{p9_11}
ds^2=(1-\frac{r_g}r)dt^2-\frac{dr^2}{1-\frac{r_g}r}-
r^2(\sin^2\theta d\varphi^2+d\theta^2)
\end{equation}

\section{Geometrization of weak interaction in the O-theory}
Choose a state $\Psi$ for $u$ and $d$-quarks (the solution found by analogy to the lepton sector, but with the other hypercharge operator $Y$)
\begin{equation}\label{p10_2}
\Psi=\sum_{i=0}^9\alpha_iu+\sum_{i=0}^9\beta_id,\quad
Y_{u,d}=\left(\matrix{-\frac13&0\cr0&\frac23}\right)
\end{equation}
here $\alpha_i$ and $\beta_i$ are coefficients where $\Psi$ corresponds to the lagrangian of the weak interactions.

Similarly to the case of weak interactions we write the total lagrangian of the O-theory, neglecting the distinction between left and right particles ($A,B=0,1,\dots,7$):
$$\L_\o=\L_f+(\partial_a\bux\Psi\varphi-
\frac i2q^AA_a^A\bux\Psi\varphi*\Sigma^A)
*(\partial^a\Psi_\varphi+
\frac i2q^BA^{a(B)} \Sigma^B*\Psi_\varphi)$$
$$+\frac i2\overline\Psi*\gamma_a(\overrightarrow\partial^a
+\frac i2c_Aq^AA^{a(A)}\Sigma^A)*\Psi
-\frac i2\overline L*\gamma_a(\overleftarrow\partial^a-
\frac i2c_Aq^AA^{a(A)}\Sigma^A)*\Psi$$
\begin{equation}\label{p10_3}
-\tilde h\overline \Psi*\Psi_\varphi*\Psi-\tilde h\overline
\Psi*\bux\Psi\varphi*\Psi)+m^2||\Psi_\varphi||^2-\frac f4||\Psi_\varphi||^4
\end{equation}

Let $\Psi_\varphi=\Psi_{\varphi_0}$ from (\ref{p4_8}). Write some members of the lagrangian $\L_o$
$$\L_\o'=
\frac{q^{(5)2}m^2}{2f}A_a^5A^{a(5)}+\frac{q^{(6)2}m^2}{2f}A_a^6A^{a(6)}
+q^6A^{a(6)}(\kappa_3\overline u\gamma_a u+\kappa_4\overline d\gamma_ad)$$
\begin{equation}\label{p10_4}+\kappa_5(q^6A^{a(6)} +iq^5A^{a(5)})\overline u\gamma_ad+\kappa_5(q^6A^{a(6)}-iq^5A^{a(5)})\overline d\gamma_au$$
$$+f^{1256}q^{12}q^{56}A_a^1A^{a(2)}A_b^5A^{b(6)}
\end{equation}

Here the group constant of weak interactions $q^{12}=g$ could be substituted.
Thus the massive charged vector bosons $D$ and $D^*$ induced gravity, but with another constant
\begin{equation}\label{p10_5}
\kappa_S=m_S^2=gq^{56}A_a^1A^{a(2)}=gq^{56}m_W^2
\end{equation}

If $m_W=100\ Gev, m_S=1\ Gev, g=10^{-2}$, then $q^{56}=10^{-2}$! 

It appears the group constant $SU(2)\times U(1)$ and $q^{56}$ possibly are equal! Or they are, at least, close.

The similar formula holds for the gravity
\begin{equation}\label{p10_6}
\kappa_g=m_D^2=q^{56}q^{47}A_a^4A^{a(7)}=10^{-2}q^{47}m_Cm_E
\end{equation}

Assume the mass of major vector fields are equal to the Planck mass (this assumption is reasonable, since it is assumed that they form a coupled state), we obtain $q^{47}=10^2$.  

Then the solution for a stable nucleus could be regarded as a solution inside of ``black hole'' in the Schwarzchild metric. The invisibility of free quarks is explained as a solution inside of ``black hole'', that ``has no hair'' therefore it is impossible to extract any information except the information of its mass, charge and its angular momentum.

In fact, the previous arguments are not completely accurate. They refer to the case where ``black hole'' is not charged and has no rotational degrees of freedom. That is applicable to the lower states of mesons. Proton can not be regarded as Schwarzchild metric external solution, because it is a charged particle. So the proton description should use the Reissner-Nordstrom solution \cite{Hawking2}. That metric is a metric of a charged ``black hole'' with the charge of $Q$ and the gravitational mass of $M$, but with the other ``strong gravitational'' constant of $\kappa_S$:
\begin{equation}\label{p10_8}
ds^2=(1-\frac{2M}r+\frac{Q^2}{r^2})dt^2-\frac{dr^2}{1-\frac{2M}r+\frac{Q^2}{r^2}}-r^2(\sin^2\theta d\varphi^2+d\theta^2)
\end{equation}
(to be completely precise we could have used the Kerr-Newman geometry, if we had taken into account the spin and rotational degrees of freedom).

Pay attention to an interesting feature of this solution, which takes place in atoms: an increase in charge with no increase in weight leads to the impossibility of the existence of ``black hole''. However complementing the nucleus with neutrons increases the mass of the nucleus and leads to increase in its gravitational radius, and as a consequence in its stability. The relation between charge and mass of stable nucleus is
\begin{equation}\label{p10_9}
Q<M=r_g/2
\end{equation}

However we should take into account the electrical repulsion of protons inside the nucleus. If the nucleus consists of protons only then the force of electrical repulsion increases and damages the nucleus. That fact is known from nuclear physics, but the model of the O-theory is a theoretical foundation.

\section{Strong interactions in the O-theory}
The research above shows the problem of quarks not flying-off nucleus
could be solved in the framework of the O-theory as weak interaction
manifestation. At the same time the geometrical nature of the
effect, and therefore taking into account lagrangian
non-associativity, leads to additional difficulties during the model
quantization. On the other side, given the nucleus has
high energy (the energy of the same order or higher than that of
charged W-boson mass) the weak interaction current values and the
geometrical structure and therefore the non-associative members in
the theory apparently could be neglected. So, under the listed
assumptions, the field part of lagrangian appears to be associative.
Assume we neglect Higgs sector and limit our consideration by
basis solution $\Psi_0$, then the O-theory lagrangian generally becomes
associative. Therefore on the octonion algebra the convolution
(\ref{p4_2}) appears to be associative scalar product of type
(\ref{p2_11}). So we can build up special octonion symmetry group
$G_2$. Once $SU(3)\in G_2$, the colour quark group $SU_c(3)$
could be introduced to the model using standard method, i.e. fitting
coefficients $\alpha_k,\beta_k$ in (\ref{p10_2}). In this paper the
described issue is not considered.

\section{Conclution}
General framework for unified theory construction was proposed in
the paper. As it had been predicted, finally we obtained the
expression for current with coefficients, which gave us certain
freedom (e.g. we were free to variate constant $\kappa_3$ inside
certain limits etc.). That new current expression was caused by a
number of new members appeared while generalization procedure of ST
lagran\-gian. Basically, those members consist of fields
$A_\mu^A,A=5,6,7,8$ and those ones appeared from multiplets on
spinor fields of $\K$ state space construction method. Should a
general research be a start point? A physical nature of something
could be a partial representation of that something and whether
there is no interesting solution of a general problem detected then
there is no interest in the general theory itself. For that reason a
matter of principle for the author was the mathematical principles
existence detection, which principles prove proposed unified theory
physical foundation. Detailed symmetry of the theory and
quantization procedure research, from author's point of view, should
be conditioned by theory approximative schemes and could be a
subject of further articles.

During gravitation build up process there was a remarkable feature
of the developing formalism. Instead of introducing the space-time
metric to the theory as a corollary of the variational principle for
gravitational field action in terms of $R$, it was introduced as a
corollary of geometrization method. Basically, the general
relativity equations as a solution on the extremals also seems to be
possible in the framework of the developed formalism. Such
supposition is justified because the metrics is defined by
vector-potentials $A_\mu^J$, the equations for which are
Euler-Lagrange equations.

\section{Acknowledgements}
The author expresses his gratitude to the participants of Friedmann
Laboratory for Theoretical Physics seminar for profound discussions.

\end{document}